\documentclass[11pt]{article}
\usepackage{cospar}
\usepackage{url}

\usepackage{graphicx}
\usepackage[figuresright]{rotating}

\title{IMPROVING THE ANGULAR RESOLUTION OF EGRET AND NEW LIMITS ON SUPERSYMMETRIC DARK MATTER NEAR THE GALACTIC CENTER}

\author{D. Hooper \address{Department of Physics, University of Wisconsin, 
    1150 University Ave, Madison, WI 53706, USA},
	and
        B. Dingus \address{Los Alamos National Lab, MS H803 P-23, Los Alamos, NM 87545, USA}}

\begin{document}

\maketitle

\begin{abstract}
Using the EGRET data and an improved point source analysis, including an energy dependent point spread function and an unbinned maximum likelihood technique, we have been able to place considerably lower limits on the gamma ray flux from the galactic center region. We also test this method on known sources, the Crab and Vela pulsars.  In both cases, we find that our method improves the angular precision of EGRET data over the 3EG catalog.

This new limit on gamma rays from the galactic center can be used to test models of annihilating supersymmetric dark matter and galactic halo profiles. We find that the present EGRET data can limit many supersymmetric models if the density of the galactic dark matter halo is cuspy or spiked toward the galactic center. We also discuss the ability of GLAST to test these models.
\end{abstract}

\section*{INTRODUCTION}

Observations by a variety of experiments have revealed that a great deal of the mass of our universe is dark and cold. Despite this growing body of evidence, we are still ignorant of the nature of dark matter.  

One of the most promising dark matter candidates is the lightest neutralino in supersymmetric models (Ellis et al., 1984; Goldberg, 1983).  In most models, the lightest supersymmetric particle (LSP) is stable by the virtue of R-parity (Weinberg, 1982; Hall and Suzuki, 1984; Allanach et al., 1999).  Often, this particle is a neutralino, $\chi^0$, the partner of the photon, Z-boson and neutral higgs bosons.  This candidate is attractive due to the fact that it is electrically neutral, not colored and naturally has the approprite annihilation cross section and mass to provide a cosmologically interesting relic density.

Many methods have been proposed to search for evidence of supersymmetric dark matter.  These include experiments which hope to measure the recoil of dark matter particles elastically scattering off of a detector (direct searches), experiments which hope to observe the products of dark matter annihilation (indirect searches) and, of course, collider experiments. Indirect searches include searches for neutrinos from the Sun, Earth or galactic center, positrons or anti-protons from the galactic halo and gamma rays from the galactic center and halo (Bergstrom et al., 2001; Bergstrom et al., 1998).

Methods of indirect detection which involve the galactic center
depend strongly on the distribution of dark matter in the galaxy.  At this
time, there is a great deal of debate and speculation over the merits of
various galactic dark matter halo models.  Numerical simulations favor
models with strong cusps in the central region, such as the Navarro,
Frenk, White (NFW) and Moore, {\it et al.} models (Navarro et al., 1996; Navarro et al., 1997; Moore et al., 1999).  These
models predict an increasing dark matter density as one approaches the
galactic center, $\rho \propto 1/r^{\gamma}$, where $\gamma$ is 1.0 for the
NFW case and 1.5 for the Moore case.  

There have been arguments made, based on observations, in the
favor of flat density core models.  These distributions,
although possible, are probably not capable of producing observable
signals from dark matter annihilation, and are not discussed in this
talk for this reason.  

Models with strong density spikes at the center of the halo have
recently receive some attention (Bertone et al., 2002; Ullio et al., 2001; Gondolo and Silk, 1999; Merritt et al., 2002).  In these models, cuspy
halos generate spikes as a result of adiabatic acretion of matter into the
central galactic black hole.  

Finally, if halo distributions are clumpy, rather than smooth, it
would be possible that less dark matter would be present in the central
region, and the dark matter signal diminished.

In this talk, we will show results for smooth, cuspy, halo
distributions of the NFW and the Moore profiles.  These distributions (as
well as spikey models) are especially interesting to gamma ray
experiments, as they provide signals from dark matter annihilation which
appear as point sources.  The angular distribution of events is
proportional to the dark matter density squared integrated over the line
of sight.  A strongly cusped distribution produces the vast majority of
the annihilation signal in an angular region much smaller than the point
spread function of EGRET, and hence are indistinguishable from a point source.

\section*{EGRET POINT SOURCE LOCATION ANALYSIS}

EGRET, the Energetic Gamma Ray Experiment Telescope, launched on the Compton Gamma Ray Observatory in 1991, is sensitive to gamma rays in the range of approximately 30 MeV to 30 GeV. During its operation, EGRET's observations included an exposure of approximately $2 \times 10^{9}\,\rm{cm}^2 \,\rm{sec}\,$ in the direction of the galactic center.  This paper presents an analysis of only the gamma rays $>$ 1 GeV because the continuum spectrum of gamma rays from neutralino annihilation peaks at higher energies and because the point spread function of EGRET improves with energy.  

Previous searches for point sources in EGRET data, such as the 3EG catalog (Hartman et al., 1999), used a single mean point spread function for each observed gamma-ray above 1 GeV and spatially binned the data in square bins of sides 0.5 degrees. However, the EGRET point spread function significantly improves with increasing gamma-ray energy with 68\% of 1 GeV gamma-rays reconstructed within 1.3 degrees of the true direction as compared to 68\% of the 10 GeV gamma rays within 0.4 degrees.  Our analysis uses the point spread function as determined by the preflight calibration (Thompson et al., 1997) for 6 energy bins above 1 GeV, and does not degrade the reconstructed gamma-ray direction to the nearest 0.5 degree bin. We use a spatially unbinned maximum likelihood analysis to determine the best localization of a point source.  The diffuse galactic background is from the same model (Hunter et al., 1997) used for the production of the 3EG catalog.  Before addressing the galactic center region, we tested our method on well known sources such as the Vela pulsar and the Crab pulsar (see figure 1).  
 
We found the position with maximum likelihood for the EGRET source near the Crab pulsar to be l=184.52,b=-5.79.  The known location of the Crab is l=184.56, b=-5.78 which is within the 95\% confidence region as determined by of our analysis.  The 3EG catalog lists the location of this source as l=184.53, b=-5.84, with the known location well outside of the 95\% confidence contour. 

The results for the Vela pulsar are similar.  We found the maximum likelihood at l=263.53, b=-2.82 with a known location of l=263.55, b=-2.79 again within the 95\% contour. The 3rd EGRET catalog lists Vela at l=263.52, b=-2.86 and the known location is well outside of the 99\% confidence contour.
 
When our technique is applied to the galactic center region, we find a point source located at l=0.19, b=-0.08.  The galactic center is excluded as the source beyond the 99.9\% confidence level, see figure 2. The 3EG 95\% confidence region includes the galactic center as shown by the circle in figure 2. We find that if this source, modeled with a differential power law spectrum with slope determined by the maximum likelihood technique to be -2.2, is included in the background of the region, the 95\% confidence upper limit on the number of gamma rays from a point source at the galactic center is 10 to 100 (depending on the spectrum of the source). By contrast, the source at l=0.19, b=-0.08 is a bright EGRET source of 370 gamma-rays.  The identification of this source is unknown (Mayer-Hasselwander et al., 1998), but this new localization agrees well with a postulated source of inverse Compton gamma rays from the electrons which create the galactic center radio arc (Pohl, 1997). 

\begin{figure}
\includegraphics[width=170mm]{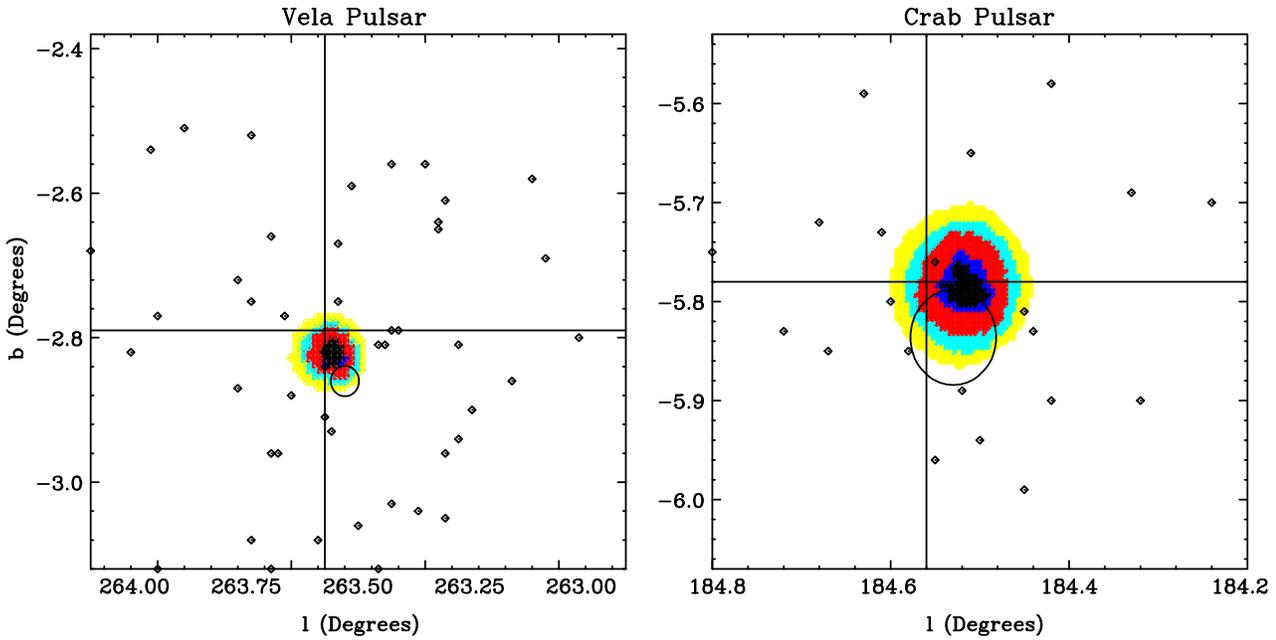}
\caption{Unbinned maximum likelihood point source analysis of the Vela (left) and Crab (right) Pulsars.  50, 68, 95, 99 and 99.9\% Confidence intervals on the point source are shown. 95\% Confidence Contour of the 3EG Catalog Positions are shown as circles for Comparison. Also Shown Are all Gamma-Rays Above 5 GeV.}
\end{figure}%

\begin{figure}
\includegraphics[width=75mm]{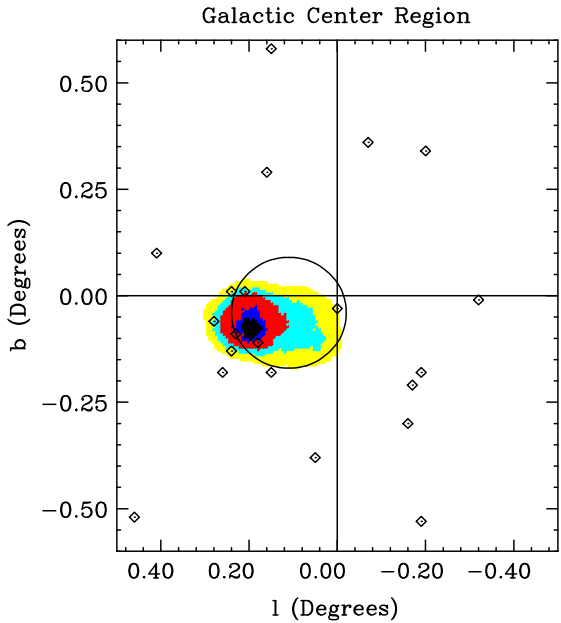}
\caption{Unbinned maximum likelihood point source analysis of the galactic center region.  50, 68, 95, 99 and 99.9\% confidence intervals on the point source position are shown. Note that the galactic center is excluded beyond the 99.9\% confidence level as the location of the source. The 95\% confidence contour of the 3EG catalog position is shown as a circle for comparison. Also shown are all gamma rays above 5 GeV.}
\end{figure}%

\section*{CALCULATING THE GAMMA-RAY FLUX FROM THE GALACTIC CENTER}

We calculated, for a variety of supersymmetric models, the number of events EGRET would have been expected to have observed, as a function of the halo model.  We only consider those models which do not violate accelerator limits, including limits on the braching ratio of b to s$\gamma$ and invisible Z decay width measurements. Furthermore, we require that the relic density of the LSP be $0.05<\Omega_{\chi} h^2<0.2$. We calculate the neutralino relic density using the full cross section, including all resonances and thresholds, and solving the Boltzmann equation numerically (Gondolo et al., 2000; Gondolo and Gelmini, 1991; Edsjo and Gondolo, 1997). Coannihilations with Charginos and Neutralinos are included. We then calculate the LSP annihilation cross section, mass and resulting gamma ray spectrum, for a given halo model. The results are shown in figure 3. 

\begin{figure}
\includegraphics[width=170mm]{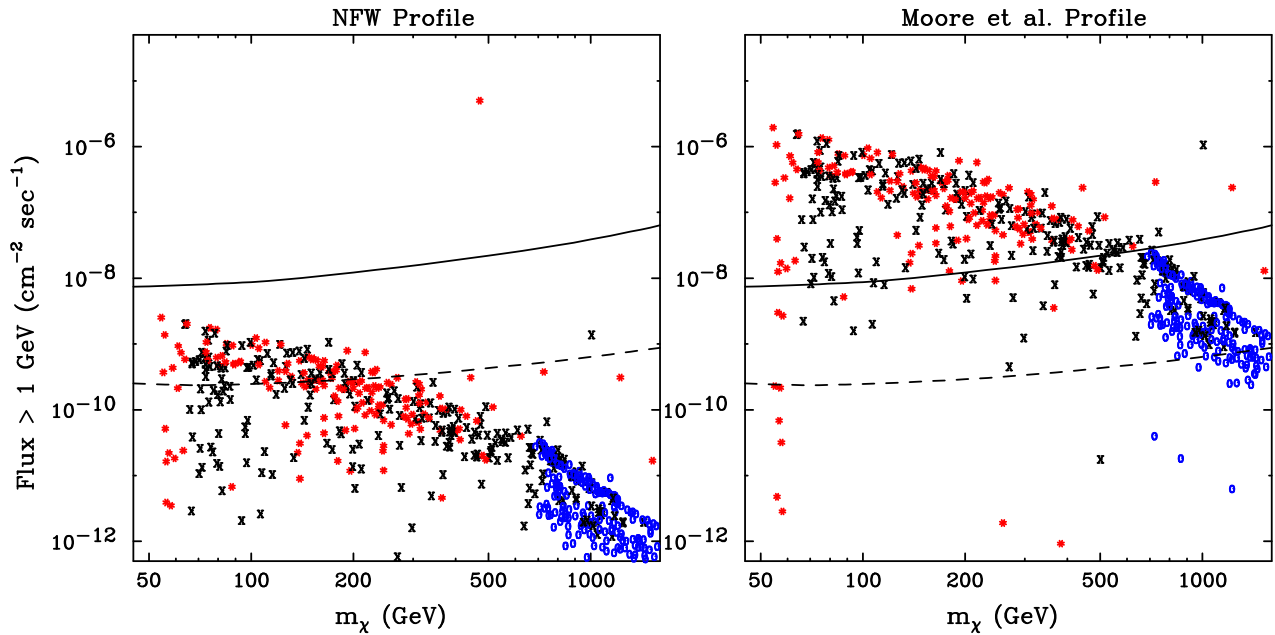}
\caption{SUSY model predicted fluxes for NFW (left) and Moore et. al. (right) halo profiles.  Also shown are the 95\% confidence upper limit of EGRET (solid line) and the expected GLAST sensitivity (dashed line). Blue (darker) circles represent models with an LSP which is more than 95\% higgsino, red (lighter) stars represent models with an LSP which is more than 95\% gaugino and black x's are models with mixed neutralinos.}
\end{figure}%

The general supersymmetric parameter space, even for the minimal supersymmetric standard model, consists of more than 100 free parameter and, therefore, must be simplifed to do any practical calculations. We considered a 7-dimensional parameter space consisting of the gaugino mass parameter, $M_2$, the non physical mass $\mu$, the ratio of higgs vacuum expectation values, $\rm{tan}\beta$, a universal SUSY mass scale, $M_{\rm{SUSY}}$, the pseudoscalar higgs mass $m_A$, and the couplings $A_t$ and $A_b$.  The gamma ray flux predicted, using a Moore et. al. halo profile, are shown as a function of supersymmetric parameter space in figure 4. The seven dimensional parameter space cannot be shown completely in such a figure, so regions shown assume $m_A=A_t=A_b=M_{\rm{susy}}=500$ GeV, where $m_A$ is the pseudoscale higgs mass, $A_t$ and $A_b$ are couplings and $M_{\rm{susy}}$ is the scale of the slepton and squark masses. Of course, other values of these parameters will yield a wider range of fluxes, as shown in figure 3.

We parametrized the continuum gamma-ray spectrum as a function of the LSP mass and calculated the 95\% exclusion confidence levels which could be placed by the EGRET data.  The parametrization depends on the neutralino annihilation branching fractions, but varies little in the majority of models.  This exclusion contour is shown in figure 3 as a solid line.  We did not consider the $\gamma \gamma$ or $\gamma Z$ line emission as these are generally above the energy range sensitive to EGRET.

We also considered the ability of the future experiment, GLAST, to
probe the galactic center for dark matter annihilations.  With
larger area and better angular resolution, GLAST, will be capable of
testing many more models than EGRET. Furthermore, GLAST, with
sensitivity to energies as high as $\sim$ 1 TeV, can test models with
heavier LSP neutralinos somewhat more easily than EGRET.  In figure3, the expected sensitivity of GLAST, after 3 years of observation, is
shown as a dashed line.

\section*{CONCLUSIONS}

Our analysis of the EGRET data in the galactic center region indicates an off-center point source, excluded beyond 99.9\% as the galactic center. Considering this source as background, we found no evidence of a point source at the galactic center and determined the 95\% confidence upper limits on the flux of gamma rays as a function of the WIMP mass.

We compared these limits to the flux predicted for a variety of supersymmetric models and galactic halo models.  We find that for very cuspy (or spikey) halo models, such as the Moore {\it et. al.} profile, the majority of viable supersymmetric models are excluded by our limit. We show that the GLAST experiment will have the sensitivity to further constrain the galactic halo profile of neutralino dark matter.

\begin{figure}
\includegraphics[width=170mm]{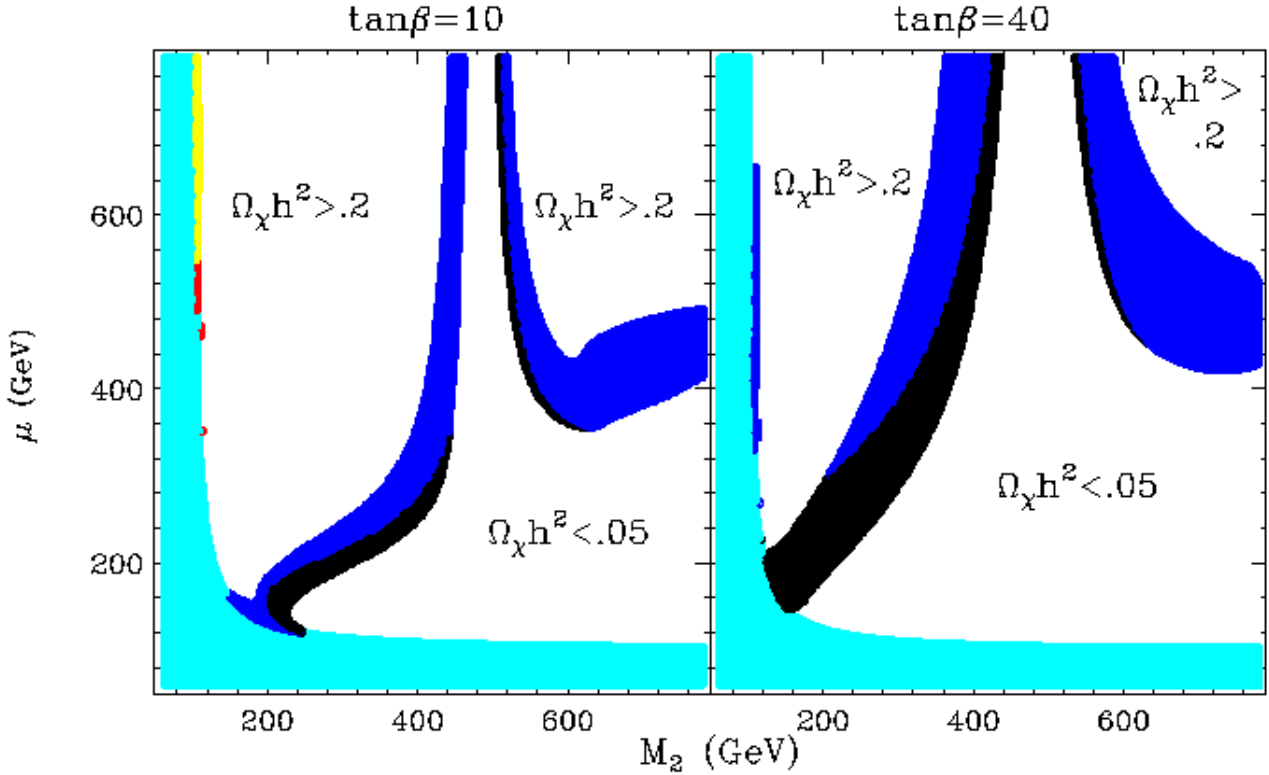}
\caption{The predicted gamma ray flux using a Moore et. al. halo profile, as calculated in figure 3, but shown as a function of supersymmetric parameter space.  From Darkest to lightest, the colors indicate a flux above 1 GeV of more than $2 \times 10^{-7}$ (black), $2 \times 10^{-8}$ (blue), $2 \times 10^{-9} \rm{cm}^{-2} \rm{sec}^{-1}$ (red), or less (yellow).  The area along the left and lower edges (cyan) is ruled out by chargino searches at LEP2. Roughly speaking, black and blue areas are excluded by our analysis (for a Moore et. al. profile). For all of the space shown here, $m_A=A_t=A_b=M_{\rm{susy}}=500$ GeV, where $m_A$ is the pseudoscale higgs mass, $A_t$ and $A_b$ are couplings and $M_{\rm{susy}}$ is the scale of the slepton and squark masses.}
\end{figure}%

\section*{ACKNOWLEDGEMENTS}

This work was based on previous work (Hooper and Dingus, 2002) supported in part by a DOE grant No. DE-FG02-95ER40896 and
NASA grant No. NAG5-9712 and in part by the Wisconsin Alumni Research Foundation.

\end{document}